\documentclass[epj]{svjour}
\usepackage{graphics}
\newcommand{\het}{($^3$He,t)}
\newcommand{\pn}{(p,n)}
\newcommand{\hetrois}{$^3$He}
\newcommand{\deu}{$^2$H}
\newcommand{\hyd}{$^1$H}
\newcommand{\car}{$^{12}$C}
\newcommand{\del}{$\Delta$}

\begin{document}
\title{ Inclusive \deu \het\ reaction at 2~GeV
 }
\subtitle{ }

\author{B. Ramstein\inst{1} \and C.A. Mosbacher\inst{2} \and
D. Bachelier\inst{1}$^\dag$  \and I. Berquist\inst{3}$^\dag$ 
\and M. Boivin\inst{4} 
\and J.L. Boyard\inst{1} \and A. Brockstedt\inst{3} \and L. Carl\'en\inst{3}
 \and R. Dahl\inst{5} \and P. Ekstr\"om\inst{3} \and C.
Ellegaard\inst{5} \and C. Gaarde\inst{5}$^\dag$ \and
T. Hennino\inst{1} \and J.C. Jourdain\inst{1} \and C. Goodman\inst{6} 
\and J.S. Larsen\inst{5} 
\and P. Radvanyi\inst{1,4} \and M. Roy-Stephan\inst{1}.}
 
\titlerunning{ \deu \het\ reaction at 2~GeV}

\institute{
 Institut de Physique Nucl\'eaire, IN2P3(CNRS), 91406 Orsay Cedex, France \and  Institut f\"ur Kernphysik, Forschungszentrum J\"ulich GmbH, D-52425
J\"ulich, 
Germany
\and Institute of Physics, University of Lund, S-223 62 Lund, Sweden
\and Laboratoire National Saturne, IN2P3(CNRS) and DSM(CEA), F-91191 
Gif-sur-Yvette Cedex, France
\and Niels Bohr Institute, University of Copenhagen, DK-2100 Copenhagen, 
Danemark
\and Indiana University, Bloomington, IN 47405, USA \\
$\dag$  deceased}
\date{Received: date / Revised version: date}
%
\abstract{
 The inclusive \deu\het\ reaction has been studied at 2 GeV for energy
 transfers up to 500 MeV and scattering angles from 0.25$^0$ up to 4$^0$.
  Data are well reproduced by a model based on a coupled-channel
  approach for describing the NN and N\del\ systems.
  The effect of  final state interaction is important in the
   low energy part of the spectra. In the delta region, the cross-section is very
  sensitive to the effects of \del-N interaction and \del N
  $\rightarrow$ NN process. The latter has also a large influence well below the
  pion threshold. 
   The calculation  underestimates the experimental cross-section between 
   the
  quasi-elastic  and the delta peaks; this is possibly 
  due to projectile excitation or purely mesonic exchange currents.

} 
\maketitle
\section{Introduction}
\label{intro}
The \het\ reaction at 2 GeV studied at Laboratoire National Saturne in both
inclusive \cite{Ellegaard83,Ellegaard85,Contardo86} and
exclusive \cite{Hennino92,Hennino93} experiments on carbon and heavier nuclei
has proven to be a very useful tool to investigate the \del-N
interaction \cite{Chanfray84,Dmitriev85}. In all charge exchange
reactions \cite{Roy-Stephan88,Gaarde91,Lind87,Prout96}, the excitation of the
\del\ resonance
involves both a spin longitudinal (pion-like) and a spin transverse
($\rho$-meson-like or photon-like) coupling, like in pion and photon induced
reactions respectively. However, in the case of charge exchange experiments,
nuclear response functions are explored in the space-like region ($\omega < $q)
that cannot be reached in pion or photon induced experiments.

Microscopic \del-hole models developed for charge exchange reactions
  suggest that the response of nuclei to the spin-isospin excitation
  induced by  charge exchange probes is especially sensitive to the
  spin-longitudinal component of the \del-N interaction, which is strong and
  attractive at the relevant momentum transfers. This results in a shift of
  the response towards lower energy transfers, in agreement with the
  observed peak positions of the \del\ resonance in the
  $^{12}$C(p,n),$^{12}$C\het\ and $^{208}$Pb\het\
  reactions \cite{Delorme91,Udagawa90}.  In
  addition, the calculations of exit channels are in qualitative
  agreement with the results of exclusive
  experiments \cite{Fernandez93,Udagawa94,Korfgen94,Kagarlis97}.  However,
  these models are not able to reproduce the whole cross-section in the
  region of excitation of the \del\ resonance and underestimate very much
  the yield in the so-called dip region lying between the quasi-elastic 
  and the \del\ peaks.

 A recent analysis within the \del-hole model of the ($\vec{p},\vec{n}$)
 polarisation observables in the \del\ region \cite{Prout96} has shown that the
 underestimate of the cross-section is mainly concentrated in the
 transverse component of the cross-section, the longitudinal contribution
 being fairly well reproduced by the model. This last result,
 together with the succesful description of coherent pion
 production data within the \del-hole model \cite{Udagawa94}, support the
 conclusion of attractive \del-hole correlations in the spin-longitudinal
 channel.

  In the quasi-elastic region, a similar result has been
  obtained \cite{Taddeucci94}. A DWIA
  calculation with a residual interaction based on a $\pi+\rho$ + g' model is
  able to reproduce the longitudinal response while the transverse response
  is underestimated by more than a factor 2.

  This excess of experimental cross-section in the transverse channel is of
  high interest. Roles of 2p-2h correlations and  meson-exchange
  currents are invoked to explain this effect, but more theoretical work is
  needed to confirm these hypotheses.

   Experimental data on the \deu\ nucleus can be helpful for understanding the
    roles of the  \del-N interaction and 
   meson-exchange currents alone. In addition, deuterium is as a two-body 
   system with a well-known wave
   function and the analysis is therefore simpler than for heavier
   nuclei. Very recently, Ch. Mosbacher and F. Osterfeld \cite{Mosbacher97}
   proposed a theoretical calculation of the \deu(p,n) reaction based on a
   coupled channel approach to describe the intermediate \del-N or NN system
   in a non-relativistic framework. This model allows a calculation of the
   energy transfer spectra in the quasielastic, dip and \del\ regions as
   well as in the different exit channels ($\pi$d,$\pi$NN,NN) and has been
   compared to the LAMPF \deu(p,n) \cite{Mosbacher97} data covering the whole
   energy transfer range up to 500 MeV and to the \deu($\vec{p},\vec{n})$
   data measured in the dip and delta regions \cite{Prout94}.  An overall
   successful description of the total energy transfer spectra is obtained
   in the quasi-elastic and \del\ regions. As in the case of \car , the
   model fails however to describe the dip region and the low energy side of
   the resonance and this discrepancy again
   arises  from the transverse component. This is interpreted by the
   authors as an effect of two-body meson-exchange currents and put together
   with the significant effect of such processes in (e,e')
   reactions \cite{Boffi93}.

    On the other hand, a small contribution from excitation of a \del\
    resonance in the projectile is also expected 
    on the low energy side of the \del\ and the exact size of this contribution
    is a subject of investigation \cite{Jo96}. 

     Previous studies of the \het\ reaction have proven that it behaves at a
     given four-momentum transfer exactly like the (p,n) reaction except for
     the \het\ form factor which produces a much steeper decrea\-se of the
     cross-sec\-tion as a func\-tion of  momentum transfer \cite{Ellegaard85}.

     No data on spin-observables have been obtained for the \deu\het\
     experiment. However, in addition to inclusive spectra, decay channels
     have been measured at La\-boratoire National Saturne and can possibly
     bring some constraints on effects of meson exchange currents and
     projectile excitation, since the former only contributes to the 2p and
     the latter to the $\pi$NN or $\pi$d exit channels. It is therefore of
     high inte\-rest to also analyse the \deu\het\ reaction in the framework
     of ref. \cite{Mosbacher97}.

    We focus in this paper on the inclusive \deu\het\ data at 2 GeV.
     After a short description of the ex\-pe\-ri\-mental
      set-up (Sec.~\ref{exp}) , we give our  experimen\-tal re\-sults
      (Sec.~\ref{data}), we present the main ingredients of the mo\-del 
      (Sec.~\ref{theory}),
       compare  cal\-culations with expe\-ri\-men\-tal
      da\-ta (Sec.~\ref{discussion})  and give our conclusions (Sec.~\ref
      {conclusion}).

\section{Experimental set-up}
\label{exp}
The \het\ experiment has been performed  using  a 2 GeV
\hetrois\ beam delivered by the synchrotron of the Laboratoire
National Saturne at Saclay. Outgoing tritons were momentum analysed with
 the Spes4 spectrometer, a D5Q6S2 instrument 35m long from target to focal
 plane \cite{Grorud81}. The maximum rigidity of this spectrometer was 4 GeV/c
 and its momentum acceptance was $\Delta$p/p$\sim$ 7.10$^{-2}$. Angular
 acceptance resulted from the combination of the beam emittance of 0.3$^0$
 (FWHM) in the horizontal plane and 1.3$^0$ (FWHM) in the vertical plane and
of the spectrometer front collimator aperture angles of $\pm 0.15^0$ in the
 horizontal plane and $\pm 0.3^0$ in the vertical plane. As detailed characteristics of the Spes4
 spectrometer and its detection system have been given  in
 references \cite{Grorud81} and \cite{Bedjidian87}, we will present only
 their main features.

 The focal plane was equipped with two drift chambers 1m apart in order to
  ray-trace scattered particles. These 1m wide and 0.2m high chambers
  consisted of three planes each, two of them with vertical wires and one 
  with wires at 45$^0$. The cell size was 5 cm for each plane and the anode
  wires were arranged in doublets in order to solve the left-right
  ambiguity. A position resolution of .5 mm (FWHM) was obtained.

   The overall resolution on the triton momentum was about
   10$^{-3}$ (i.e. 3 MeV at 2 GeV) and was dominated by beam spot size
   on target. An overall efficiency of about 95$\%$ was obtained for the
   measurement of the tritons.

   To cover the quasi-elastic and \del\ peaks, tritons with energies between
   1.5 and 2 GeV were detected. The whole spectrum was obtained by means of
   four field settings of the spectrometer. The  70-100 MeV wide overlap
   regions were in satisfactory agreement. 
   
  The data taking was triggered by a coincidence signal between two
  scintillator hodoscopes situated 16 meters apart. Due to the
  small momentum acceptance of the spectrometer, only tritons reached the
  detectors, except at the lowest magnetic fields, where a small
 deuteron  contribution was rejected using 
  a time of flight measurement between the two hodoscope planes.

 The \deu\het\ reaction has been studied at 0.25$^0$, 1.6$^0$, 2.7$^0$
 and 4.0$^0$. The SPES4 spec\-trometer is not  movable and the dif\-ferent scattering
 angles were achieved with a beam swinger device aimed at changing the
 angle of the beam impinging on the target. For each measurement, the
 scattering angle was deduced from the positions of the beam on two
 insertable wire chambers located in the target area. The residual angular
 off-set was obtained with a precision of 0.07$^0$ (rms) by measuring the
 cross-section of the Gamow-Teller peak in the reaction
 $^{12}$C($^3$He,t)$^{12}$N$_{gs}$ for several angular settings of the
 spectrometer, both right and  left of the beam direction.\par

  The cross-sections on \deu \ have been obtained by subtraction of the
 yields obtained on CD2 and C targets. Beam intensities of about
 10$^9$-10$^{10}$ particles/s and target thicknesses of 200 mg/cm$^2$ were
 used. The average contribution from the carbon nuclei to the CD2 target
 is about 30 to 45 $\%$ depending on the angle. The discrete states of
 the $^{12}$C nucleus that show up as narrow peaks in the CD2 spectrum at
 the smallest angles allow to check the
 validity of the subtraction with good precision.
\par

 As described in ref. \cite{Bergqvist87}, the  cross-section absolute 
 normalisation was obtained using the known elastic cross-section of $^3$He
 on protons. Due to the uncertainties in these cross-sections and in target
 thicknesses this  overall absolute normalisation was determined within 
  15$\%$.

\section{Experimental results}
\label{data}
\begin{figure}

\resizebox{0.50\textwidth}{!}{
  \includegraphics{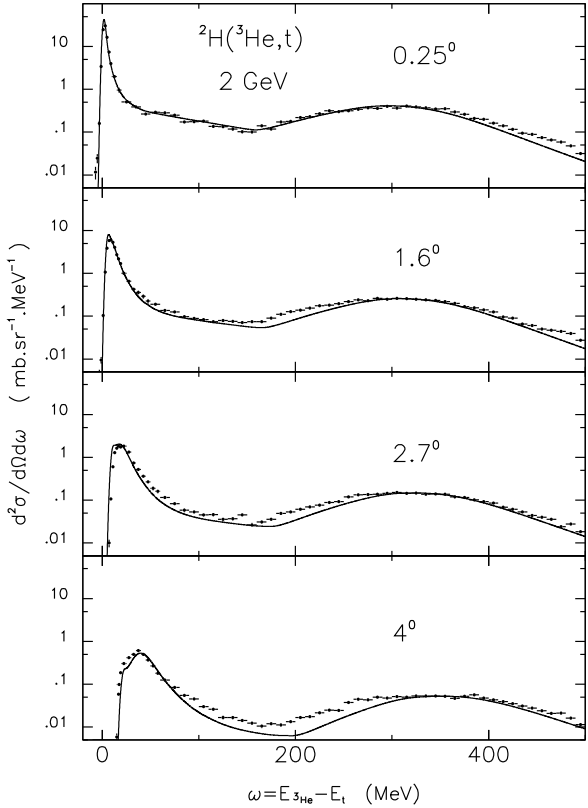}
}
\caption{
 Energy transfer spectra (full dots) at  4 angles on a
logarithmic scale compared to the calculation described
in section IV (full line). The calculations have been folded with the
experimental energy resolution and angular acceptance.
 }
\label{fig4anglog}       
\end{figure}
\begin{figure}

\resizebox{0.50\textwidth}{!}{
  \includegraphics{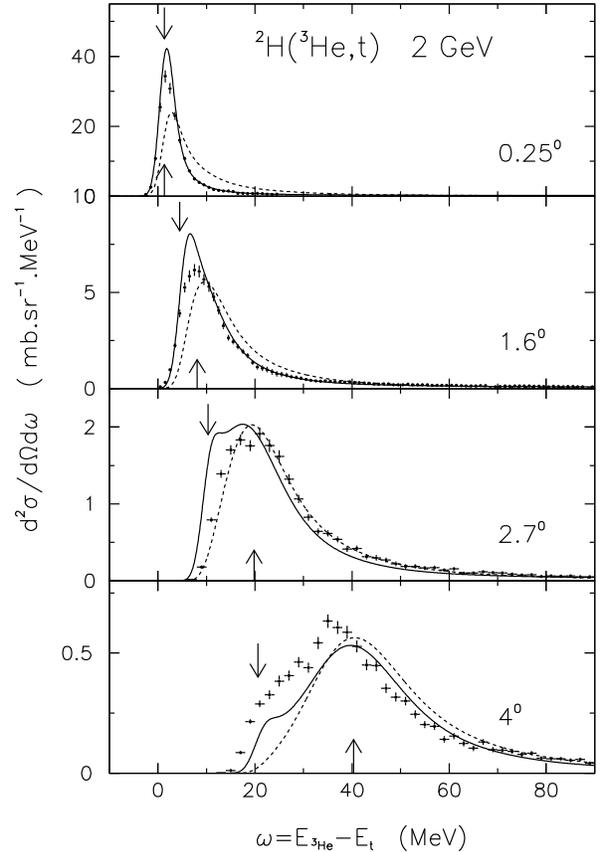}
}
\caption{ Energy transfer spectra in the low-energy region (full dots) compared
 to the full calculation (full line) and to the calculation without FSI
 (dashed line). The calculations have been folded with the
experimental energy resolution and angular acceptance.
 The upper and lower arrows indicate respectively the energy
 transfers for kinematics on a 2 proton pair with zero relative momentum
 and quasi-free kinematics on a nucleon at rest. These values are
 calculated  for the mean angles accepted
 by the set-up(see text).
 }
\label{fig4angqf}       
\end{figure}
\begin{figure}

\resizebox{0.50\textwidth}{!}{
  \includegraphics{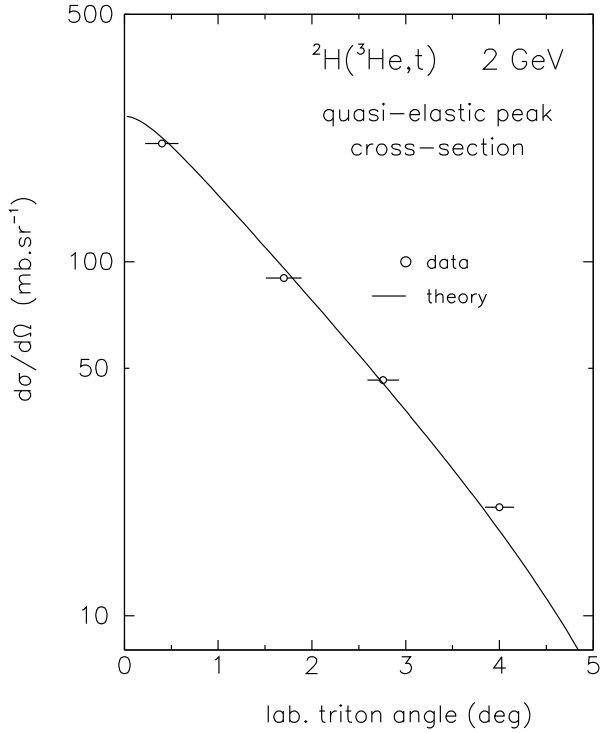}
}
\caption{ Angular distribution of the quasi-elastic  peak (open circles)
compared to the theoretical prediction. The abscissa and the horizontal bar of
each data point take into account angular acceptance effects.}
\label{distangqf}       
\end{figure}
\begin{figure}
\resizebox{0.50\textwidth}{!}{
  \includegraphics{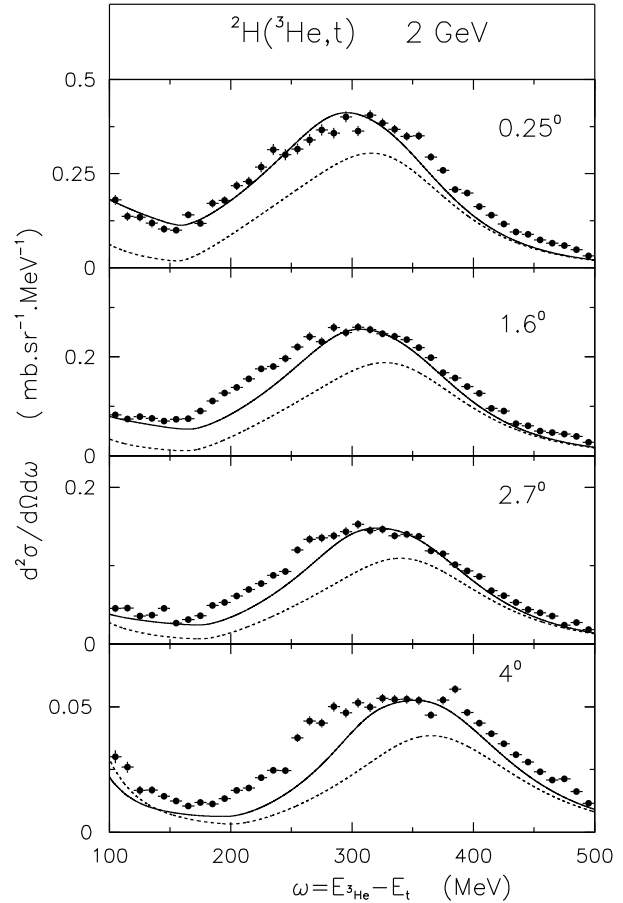}
}
\caption{ Energy transfer spectra in the \del\ region (full dots) compared to
the complete calculation (full line) and  to the spectator
approximation calculation, i.e. without $\Delta$-N
interaction nor \del N $\rightarrow$ NN process (dotted line).
 The calculations have been folded with the
experimental energy resolution and angular acceptance.
}
\label{fig4angdel}       
\end{figure}
\begin{figure}

\resizebox{0.50\textwidth}{!}{
  \includegraphics{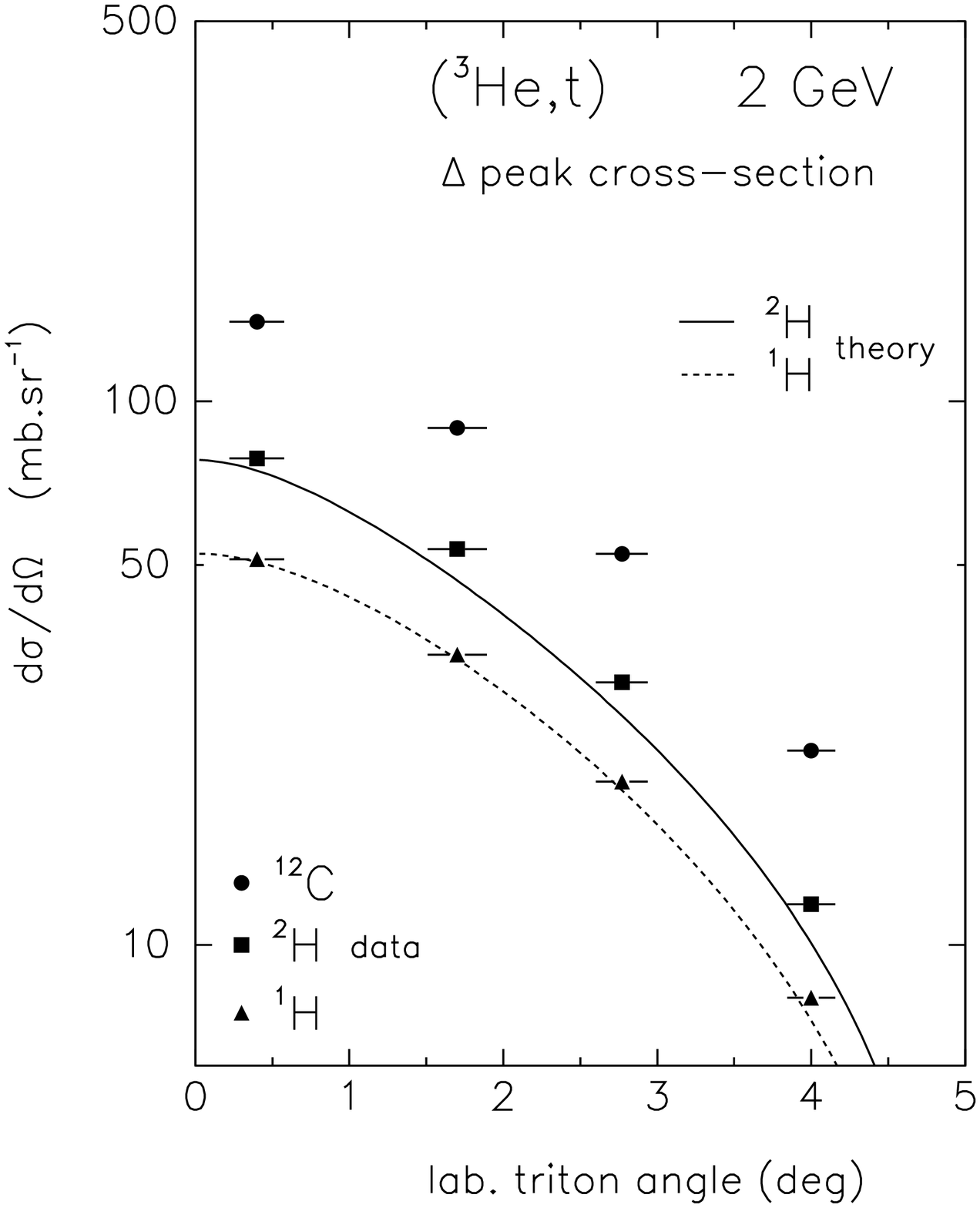}
}
\caption{ Angular distribution of the \del\ peak for the \deu \het\ reaction
compared to the results of ref. \cite{Contardo86} on \hyd\ and \car\
targets.  The abscissa and the horizontal bar of
each data point take into account angular acceptance effects.
 The theoretical predictions for the \deu\ and \hyd\ 
nuclei are shown as  full and  dashed line respectively.
}
\label{distangdel}       
\end{figure}

The whole energy transfer distributions obtained in our experiment at four
   different angles of the spectrometer are displayed on
   fig~\ref{fig4anglog} on a logarithmic scale. Two structures show up very
   clearly: a quite narrow peak at low energy transfers corresponding to
   quasi-elastic mechanisms involving only nucleonic degrees of freedom and
   a broad bump, above the pion threshold, corresponding to the excitation
   of a nucleon into a \del\ resonance. Figs~\ref{fig4angqf} and
   \ref{fig4angdel} show more precisely these two peaks on a linear scale.
   We have arbitrarily set the dividing line between the two contributions
   at an energy transfer of 140 MeV, and the respective yields have been
   integrated and are plotted on figs.~\ref{distangqf} and \ref{distangdel}.
   For each measurement, the distributions of scattering angles accepted by
   the set-up have been calculated, taking into account beam emittance and
   collimator aperture. The angles and the horizontal bars reported on the
   figures correspond respectively to the mean values and the rms of these
   distributions. The angle mean values are 0.40$^0$, 1.70$^0$, 2.76$^0$ and
   4.0$^0$ when the spectrometer is set at 0.25$^0$, 1.6$^0$, 2.7$^0$ and
   4$^0$ respectively. The rms angular acceptance is about 0.17$^0$ for the
   four settings.


    The integrated cross-section of the low energy peak decreases by a
factor 11 between 0.25$^0$ and 4$^0$ as can be seen from the figures
\ref{fig4anglog} to \ref{distangdel}
but the most impressive effects are the shift  of the peak position and
the increase of its width as the angle gets larger. The \del\ excitation
cross-section has a smoother behaviour as a function of angle since the
cross-section only decreases by a factor 7 from 0.25$^0$ to 4$^0$ and the
width stays about constant. However, the position of the maximum shifts by
about 45 MeV from 0.25$^0$ to 4$^0$ .

\begin{figure}

\resizebox{0.50\textwidth}{!}{
  \includegraphics{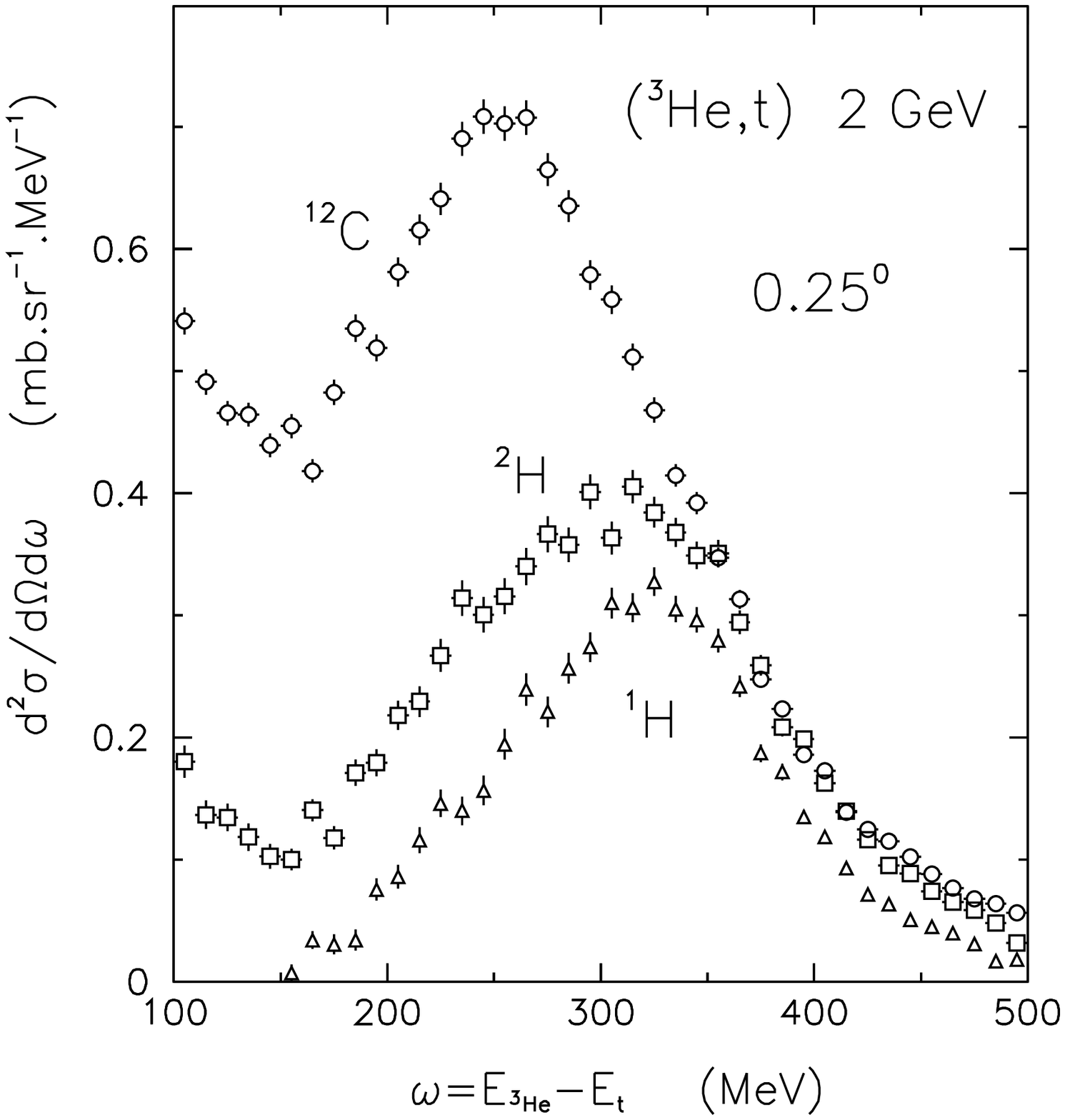}
}
\caption{ Energy transfer spectra in the \del\ region for the \deu \het reaction 
compared to the results of ref \cite{Contardo86} on \hyd\ and \car\  nuclei.
The calculations have been folded with the
experimental energy resolution and angular acceptance.
}
\label{fig3cib0deg}       
\end{figure}

The presence of these two structures and their different behaviour as a
      function of angle is a very general feature of all charge exchange
      reactions \cite{Ellegaard87,Gaarde91}. For the smallest angles in
      heavy target nuclei, the low energy peak is due mainly to spin-isospin
      excitations leading to the ground state, excited states or giant
      resonances of the resulting (Z+1,N-1)
      nucleus \cite{Bergqvist87,Brockstedt91}. Quasi-elastic processes
      corresponding to reactions on a quasi-free nucleon are quenched by
      Pauli blocking for the smallest momentum transfers but dominate the
      spectra at larger angles.

      For \het\ on a deuterium nucleus, final state interaction(FSI) will modify the
     spectrum in the quasi-elastic region. In particular, there is a strong
     effect of the $^1$S$_0$ quasi-bound state for the smallest relative
     momenta between the 2 protons, ( i.e. the smallest momentum transfers).
     A sizeable effect of this final state interaction in the $^2$H(p,n)pp
     reaction at 800 MeV and 1 GeV has been previously found
     \cite{Aladashvili77,Sakai87,Deloff93,Itabashi94}, so that we can expect
     this effect to contribute also in the case of the \het\ reaction. As an
     indication, energy transfers calculated in quasi-free kinematics and
     with kinematics corresponding to 2 protons with no relative energy have
     been indicated on fig.~\ref{fig4angqf}. The energy transfer regions,
     where the quasi-free and the quasi-bound $^1S_0$ final state are
     expected to contribute, separate when the angle increases.

  It was shown in ref.~\cite{Contardo86} that the \del\ excitation
      cross-section in nuclei was following a universal trend as a function
      of angle. This is also the case for the \deu\ target as shown on
      fig.~\ref{distangdel}, where the angular distribution on \deu\ is
      compared to the ones measured on \hyd\ and \car\ \cite{Contardo86}. It
      shows that the shape of the angular distribution is mainly dominated
      by \het\ form factor effects, whereas the position of the maximum of
      the $\Delta$ resonance depends on the target (see fig.~\ref{fig3cib0deg}) and is thus
      sensitive to the presence of other target nucleons.

      The \del\ peak appears about 25 MeV lower on the \deu\ target than on
      \hyd\ at 0.25$^0$. For \car\ and heavier nuclei, the universal 70 MeV
      shift towards lower energy with respect to the \hyd\ target has been
      stressed for a long time. It has been shown that the residual
      $\Delta$-N interaction was responsible for a 25-30 MeV
      shift \cite{Udagawa90,Delorme91}, the other half being due to more
      conventionnal effects, such as the absorption of the $\Delta$
      resonance ($\Delta$N$\rightarrow$NN) and the combined effects of Fermi
      broadening and of the steep \het\ form factor. In this respect, the
      specific interest of the deuterium nucleus is to allow a direct study
      of the effect of the $\Delta$-N interaction, the Fermi motion effects
      being exactly treated. These suitable properties of the deuterium
      nucleus have been well exploited in the theoretical interpretation of
      pion and photon induced
      reactions \cite{Maxwell80,Niskanen95,Wilhelm96,Kamalov97,Pena92,Ericson88,Garcilazo90}
      and in the recent calculation of the \deu (p,n) reaction by Ch.
      Mosbacher and F. Osterfeld, which we will adapt to describe the \het\
      experiment.

\section{Theoretical framework}
\label{theory}

\begin{figure}
\resizebox{0.50\textwidth}{!}{%
  \includegraphics{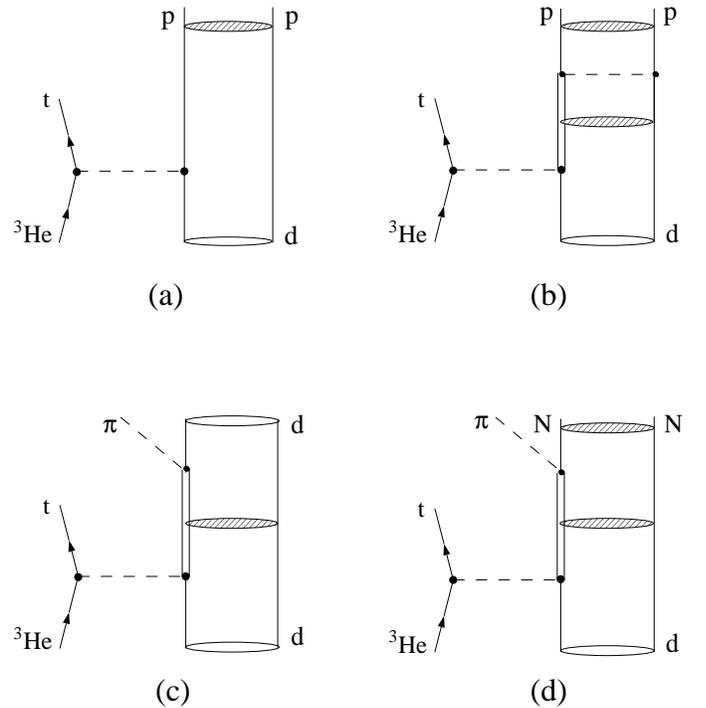}
}
\caption{ Feynman diagrams for the reaction mechanisms included in the
     model of ref.\ \protect \cite{Mosbacher97}.
      They represent:
     (a) quasi--elastic scattering, (b) \del-N$\rightarrow$ NN process, 
     (c) coherent pion production, (d) quasifree $\Delta$ decay.
     The shaded areas indicate the intermediate $\Delta N$ 
     and  $NN$ final state interaction.
}
\label{diagrams}       
\end{figure}
The model of \cite{Mosbacher97} is most suitable to describe our data,
since it allows a description of the charge-exchange reaction in the 
full energy transfer range covered by our experiment, that means 
including both quasi--elastic and $\Delta$ excitation processes.

The scattering mechanisms considered in the model are represented
by the diagrams of fig.~\ref{diagrams}. The theoretical framework is
based on a coupled channel formalism which allows one to include   both
the intermediate $\Delta $N interaction and the NN final state 
interaction in infinite order. The corresponding interaction potentials 
are constructed in a meson exchange model \cite{Machleidt87}
where $\pi$, $\rho$, $\omega$ and $\sigma$ exchange are taken 
into account. The $\Delta$ resonance is treated thereby as a 
quasi--particle with a given mass and an intrinsic energy--dependent 
width. Evaluation of matrix elements involves the propagation
of correlated two particle systems, the wave functions of which
are calculated in configuration space using the source function 
formalism and the Lanczos method. For a more detailed explanation,
please see refs.\ \cite{Mosbacher97,Udagawa94}.

Inputs of the model are the meson and baryon masses, coupling 
constants, and form factor cutoffs as given in \cite{Mosbacher97} (Table I). 
All of them are fixed by the $^2$H$(p,n)$ data and by other sources 
(such as pion absorption on the deuteron, and $NN$ scattering, e.g.). 
Furthermore, effective parameterizations of $NN \! \to \! NN$ and 
$NN \! \to \! N \Delta$ transition matrices are used in order to describe
the quasielastic and the $\Delta$ excitation in the deuteron caused by
projectile. 

With respect to the $(p,n)$ reaction, the transition
matrices for the ($^3$He,t) reaction are modified in two aspects. 
First, a ($^3$He,t) transition form factor has to be applied because of the 
spatial extension of the projectile--ejectile system. In the present work, 
we use the form factor of Desgrolard et al.\ \cite{Desgrolard92} 
which is based on a three--body Fadeev calculation. 
Second, an additional vertex factor $Z(s_\Delta,t)$ was introduced 
in order to account for off--shell corrections in the case of the 
$\Delta$ excitation. Here, we follow the arguments of Dmitriev et al.
and  adopt their specific choice for 
the $Z(s_\Delta,t)$ vertex factor, see eq.\ (12) in \cite{Dmitriev86}. 

\begin{figure}
\resizebox{0.50\textwidth}{!}{
  \includegraphics{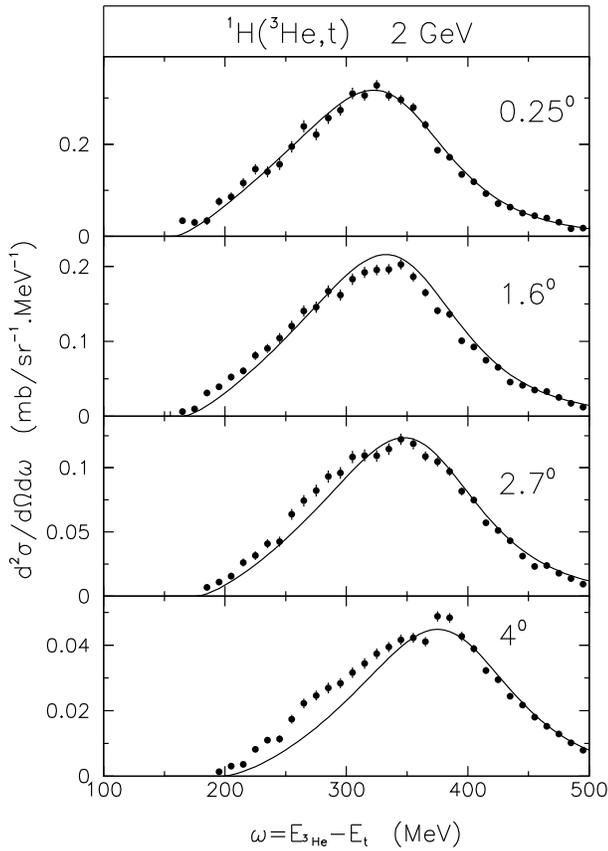}
}
\caption{  Comparison of the calculation (full line) with the experimental data
in  \hyd\het reaction at 2 GeV. The calculations have been folded with the
experimental energy resolution and angular acceptance.
}
\label{fighyd}       
\end{figure}
With the two modifications mentioned, we achieve a very good description
of the $^1$H($^3$He,t)$\Delta^{++}$ charge exchange reaction at 2 GeV.
The quality of the fit is demonstrated in fig.~\ref{fighyd}, where experimental 
$^1$H($^3$He,t) data are compared to the theoretical cross sections. 
Obviously, the   energy transfer  and 
scattering angle dependence can be both reproduced correctly. As compared
to the parameterization given in \cite{Udagawa94}, the different 
($^3$He,t) transition form factor and the additional vertex factor 
$Z(s_\Delta,t)$ clearly improve the description at non--zero scattering
angles and for low energy transfers, respectively. At the same time,
the model preserves full consistency with the $(p,n)$ charge exchange
reaction data on the proton as well as on the deuteron target.

\section{Discussion}
\label{discussion}
Results of the full calculation described in section 4 are compared to
 our
\deu\het\ experimental results  on figures \ref{fig4anglog} to
\ref{distangdel}.
An overall good agreement is observed for the whole spectrum at the four
angles. Both the quasi-elastic and the \del\ peak cross-sections are
satisfactorily described in shape and magnitude. For the low-energy peak,
the very fast evolution of the cross-section with angle is well reproduced by
the calculation.

To achieve a consistent comparison with the experimental energy transfer
spectra, the calculation has been folded with a gaussian ($\sigma$=1.3 MeV)
to account for  the experimental energy resolution. This is necessary in the
quasielastic region for the smallest angles where the theoretical peak is
very narrow.

 Concerning the effect of angular resolution, we weigh\-ted the theoretical
angular distribution by the angular transmission of the set-up in order to
take into account both collimator aperture and  beam emittance in the
theoretical spectrum and make the comparison as fair as possible.

A limitation of our experiment arises from the fact that the
exact angle of the measurement is known to an overall offset of
$\pm$0.07$^0$ (rms). Since the peak position  and width  vary as the
square of the scattering angle and the slope of the cross-section also
increases with angle, the sensitivity of the energy transfer spectra to
this offset is the largest at 4$^0$. The biggest effect that can be
expected at this angle is a shift of the low energy side and of the peak
position of the spectrum by about $\pm$ 2 MeV, together with a rescaling of
the maximum cross-section of about $\pm 10\%$. The high energy side of the
spectrum is insensitive to such variations of angle. Such small effects don't
hinder the comparison of the theory to the experiment. 

On fig.~\ref{fig4angqf}, we clearly see the important role played by the
final state interaction in the model. With respect to the calculation in the
spectator approximation, the yield is concentrated at smaller energy
transfers. For the largest angles, a shoulder is visible in the theoretical
curve at an energy
transfer corresponding to small relative momenta of the 2 protons, due to
the interaction in the $^1S_0$ partial wave. The inclusion of 2p final state interaction improves the agreement
with experiment at 0.25$^0$, 1.6$^0$ and 4$^0$. At 2.7$^0$, the curve
without FSI provides a better description of the data, which is not 
understood so far.

\begin{figure}
\resizebox{0.50\textwidth}{!}{  \includegraphics{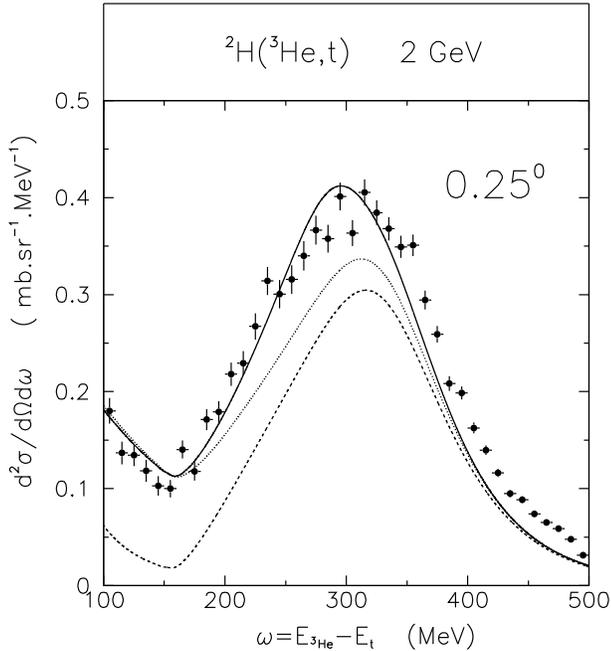}
}
\caption{ Energy transfer spectra (full dots) at 0.25$^0$ compared to the
complete calculation (full line) and to the calculations without \del-N
interaction (dotted line) and in the spectator approximation i.e. without
$\Delta$-N interaction nor \del N $\rightarrow$ NN process (dashed
line).
}
\label{figvdn0}       
\end{figure}

On figs.~\ref{fig4angdel} and ~\ref{distangdel}, we focus on the \del\
region. The description of the data is good especially at the smallest
angles. The importance of
\del-N interaction and of the $\Delta N \rightarrow NN$ transition potential
is also illustrated on figs.~\ref{distangdel} and \ref{figvdn0}. The
calculation in the spectator approximation, that means neglecting these
interactions, underestimates the cross-section   and peaks at too high an energy
transfer. 

The effect of \del-N interaction has been studied in great detail by Ch.
 Mosbacher and F. Osterfeld, in the calculation of the \deu\pn\ reaction at
 800 MeV covering the same four-momentum transfer regions
 \cite{Mosbacher97}. They have shown that the shift of the spectrum towards
 low energy transfers induced by \del-N interaction is mainly due to its
 spin-longitudinal part which arises from the pion exchange.  This
 attraction results mainly from the interference between direct and exchange
 terms. The $\rho$ meson exchange tensor part partly cancels this
 attraction, but the final result is still a shift of the spectrum towards
 lower energy transfers.

 It is clearly seen on fig.~\ref{figvdn0} that the $\Delta N \rightarrow NN$ transition
 is responsible for a large fraction of the cross-section in the so-called
 dip region lying between the quasi-elastic  and the \del\ peaks, but it
 contributes also to the enhancement and shift of the cross-section in 
 the \del\ resonance region.

The cross-section in the dip region and on the low energy side of the
resonance is well reproduced at 0.25$^0$ but an increasing underestimate by
the model is observed as the angle increases. This effect was even worse in
the case of the (p,n) reaction and was shown to be due to the
transverse contribution of the cross-section, the longitudinal one being
well reproduced. 

As mentioned in ref.~\cite{Mosbacher97}, the origin of this deviation 
may be due to purely mesonic exchange currents, which are
 not included in the present model. The contribution of these processes in the
 spin-transverse channel has been estimated in the case of the \deu(p,n)
 reaction at 800 MeV and was found to contribute significantly 
 in the dip region \cite{Mosbacher98,Jo96}.

 However, we also observe a small discrepancy on the low-energy side of the
 \del\ peak at the higher angles in the case of the reaction on the proton 
 and the question
 arises whether this could be due either to   \del\ excitation  in the
 projectile or to non-resonant pion production, none of which was included
 in the present calculation.  Both contributions are expected to contribute
 mainly on the low energy side of the resonance and projectile excitation is
 expected to be the more important \cite{Oset89,Jo96}. Our
 simulations of the projectile excitation process show a
 significantly broader angular distribution than in the case of target
 excitation. These two qualitative arguments favour the interpretation of
 the residual discrepancy in terms of projectile excitation. Furthermore, 
 the fact that this discrepancy is slightly worse in the case of \deu\
 than in the case of \hyd\ goes also in the right direction. It has indeed
 been stressed for a long time that the relative weight of the projectile
 excitation was expected to be enhanced by a factor 3 in the deuterium
 nucleus with respect to the proton, due to isospin
 coefficients \cite{Oset89}. However, our analysis shows that the
 contribution of the projectile excitation to the energy transfer spectrum
 is quite small and that it is not responsible for the shift observed from
 \hyd\ to \deu\ nuclei.

\section{Conclusion}
\label{conclusion}
We have presented data obtained in the \deu\het\ reaction at 2 GeV at
 0.25$^0$,1.6$^0$,2.7$^0$ and 4$^0$ for energy transfers ranging from the
  quasi-elastic region up to the \del\ resonance region. 
  
  The analysis is
  performed in the framework of a model derived from \cite{Mosbacher97} and
  based on a coupled channel approach to describe the NN and \del-N systems.
  This model has been used to calculate the \deu(p,n) reaction at
  800 MeV. Taking advantage of the detailed study of the
  \hyd\het\ reaction, we use in this work a slightly different
  parametrisation of the \del\ resonance \cite{Dmitriev86} excitation process 
  and introduce the
  \het\ form factor of ref.~\cite{Desgrolard92}. 
  
  The model reproduces very well the quasielastic  and \del\ peaks.
    The FSI between the 2
  protons is shown to modify significantly the spectrum in the low energy region at
  the smallest angles. In the \del\ region, both the $\Delta N \rightarrow NN$
  transition and the $\Delta-N$ residual interaction increase and shift the
  cross-section towards lower energy transfers. The effect of the $\Delta N 
  \rightarrow NN$
  transition well below the \del\ resonance is also clearly demonstrated.
  The whole spectrum is very
  well reproduced for the smallest angle. However, at large angles, an
  increasing underestimate of the cross-section in the dip region and on
  the low energy side of the resonance is observed. These small deviations
  from the model  have to be put
  together with the excess of cross-section observed in the transverse channel
  in the \deu(p,n) reaction and can  possibly
  be ascribed to projectile excitation or  meson exchange currents which are
  not included in the model.

    We plan to extend this work to the analysis of the exclusive \deu\het\
    measurements studied at 2 GeV at Laboratoire National Saturne.
    Measurements of the decay channels offer a chance to confront the model
    in a more selective way and possibly to understand the deviations
    observed in the dip region and on the low energy side of the resonance.

\section{Acknowledgements}
We are very grateful to our colleagues M. Bedjidian, D. Contardo, J.Y.
Grossiord, A. Guichard, R. Haroutunian and J.R. Pizzi from Institut de
Physique Nucl\'eaire de Lyon for their friendly and fruitful collaboration
to the experiment. We thank the staff of
Laboratoire National Saturne for having provided a very steady beam and for 
their effective technical support. 

%

%
%
%
%

%
%

\end{document}